\documentclass[aps,twocolumn,groupedaddress,amssymb]{revtex4}

\usepackage{latexsym}
\usepackage{amsbsy}
\usepackage{amsmath}
\usepackage{graphicx}
\usepackage{xcolor}
\usepackage{hyperref} 
\usepackage[capitalise]{cleveref} 
\hypersetup{
    colorlinks=true,
    linkcolor=black,
    urlcolor=blue,
    citecolor=blue
}
\usepackage{nameref}  

\begin{document}

\author{Shreyashi Sinha, Shantanu Pathak, Saswata Bhattacharya and Sujit Manna\footnote{Contact author: smanna@physics.iitd.ac.in}}
\affiliation{Department of Physics, {Indian Institute of Technology Delhi}, Hauz Khas, New Delhi 110016, India}

\preprint{APS/123-QED}

\title{Imaging Quantum Well States of Dirac Electrons in Exfoliated 3D Topological Insulators}

\date{\today}

\begin{abstract}

We present a controlled mechanical exfoliation technique for bulk 3D topological insulators that yields atomically clean ultrathin flakes, enabling quantum well states (QWS) of Dirac electrons to be clearly resolved. Achieving reliable fabrication of pristine, high-quality two-dimensional layers suitable for atomic-scale spectroscopy remains a central experimental challenge in uncovering their emergent quantum states and realizing device-relevant functionalities. Atomically resolved scanning probe microscopy and micro-Raman spectroscopy reveal a strong correlation between Raman intensity and film thickness, enabling rapid identification of (Bi\textsubscript{0.1}Sb\textsubscript{0.9})\textsubscript{2}Te\textsubscript{3} flakes with desired thickness. High resolution scanning tunneling spectroscopy on exfoliated flakes with atomically flat terraces reveals QWS, driven by quantum confinement of Dirac electrons. This effect is rarely observed due to the electrons resistance to electrostatic confinement caused by Klein tunneling. The standard phase accumulation model accurately captures the characteristics of QWS and extracts the electronic band dispersion, showing excellent agreement with density functional theory calculations. Band structure calculation reveals that with increasing quantum-layer thickness, the interlayer coupling enhances the electronic dispersion, progressively reducing subband splitting and giving rise to bulk-like continuous bands. Spatially resolved spectroscopy around surface defects further confirms that QWS of Dirac electrons in topological insulators remains robust against defect scattering. This work paves the way for exploring diverse quantum phenomena and device applications through quantum confinement, surface-state engineering, and tunable topological phases.
\end{abstract}
\maketitle

\section{Introduction} 

Superlattice engineering and quantum confinement in low-dimensional systems play a pivotal role in condensed matter physics, enabling deliberate control of electronic band structures and facilitating the discovery of novel quantum phenomena \cite{chaves2020bandgap, krizman2018tunable}. Recent breakthroughs in Moiré superlattices of graphene and transition metal dichalcogenides (TMDs) have demonstrated that subtle control of layer orientation and thickness can induce correlated insulating states, unconventional superconductivity, and topologically nontrivial phases \cite{cao2018unconventional,  andrei2020graphene, tilak2022moire}. Extending such approaches to topological insulators (TIs) is an exciting frontier \cite{rachel2018interacting,matsugatani2018connecting,wang2015low,klaassen2025realization}. The unique electronic structure of low dimensional TIs, an insulating bulk with symmetry protected boundary states, makes them promising for future quantum technologies \cite{bernevig2006quantum,hasan2010colloquium,konig2007quantum,zhao2020strain,sajadi2018gate}. In 3D bulk TIs, the bulk conduction often masks the signature of isolated surface states due to imperfect insulating behavior. Exfoliation to few-layer may suppress the bulk contribution, allowing better isolation of Dirac surface states. Furthermore, exfoliating a 3D TI bulk sample into thin layers is crucial for advancing artificial heterostructure engineering. Early exfoliation studies by Checkelsky \textit{et al.} \cite {checkelsky2011bulk} demonstrated the feasibility of obtaining thin flakes of topological insulators through mechanical exfoliation, revealing the potential to access surface-dominated transport. However, for spectroscopic measurements such as scanning tunneling spectroscopy (STS) and angle-resolved photoemission spectroscopy (ARPES), maintaining pristine, atomically flat surfaces and minimizing contamination during transfer remained major challenges. Over the past decade, substantial progress has been made in exfoliation and surface preparation techniques, particularly in the study of twisted graphene and other two-dimensional materials \cite{li2021imaging, wong2015characterization, vancso2016intrinsic, wang2022real}. Despite these advances, controlled exfoliation of topological insulators under inert or ultra-high-vacuum environments, along with improved substrate cleaning, handling protocols, and optimized post-transfer annealing procedures, remains limited to date. Further development of these methods is expected to significantly enhance the surface cleanliness and stability of exfoliated flakes. Such advancements are essential for enabling surface-sensitive probes such as scanning tunneling microscopy (STM) and spectroscopy to access the intrinsic electronic properties of topological materials with minimal extrinsic interference. This is particularly important for applications such as topological superconductors and topological magnets, including the advancement of the quantum anomalous Hall effect research \cite{yang2020multiple, li2015proximity, li2015magnetic}. Precise control over thickness and systematic investigations of quantum confinement effects in exfoliated TIs remain scarce compared to graphene and TMDs, leaving natural questions about how thickness, substrate interactions, and surface coupling influence the electronic structure in these systems.

Engineered heterostructures, such as TI-superconductor or TI-magnet interfaces, require atomic level control over thickness and stacking to tune proximity-induced effects. Controlled hybridization between top and bottom surface states in ultra-thin layers creates 2D quantum confined systems with tunable bandgaps or topological phase transitions \cite{neupane2014observation, zhang2010crossover}. A key step towards this goal is to understand how the electronic structure of TIs evolves at reduced dimensions. In ultrathin films, restricting electron motion to two dimensions causes quantum confinement, leading to discrete energy levels and significant changes in the electronic structure near the Fermi level. This results in oscillatory behavior in key physical properties such as electrical resistivity, superconducting transition temperature, surface energy, and chemical reactivity \cite{jal1988quantum, manna2013effect, guo2004superconductivity, wu2023thickness}. These quantum size effects, which exhibit atomic-layer sensitivity, are prominently observed in metallic systems like copper, Pb(111) \cite{manna2013effect, becker2010scattering}. In TIs, the quantum confinement of bulk states near the surface, induced by electrostatic band bending, leads to the formation of quantized subbands that coexist with topological surface states \cite{king2011large, bianchi2010coexistence}, resulting in more complex transport \cite{bahramy2012emergent,xia2009observation, zhang2009topological, dey2014strong, mo2016influence}. Dirac electrons in TIs exhibit spin-momentum locking, which prevents backscattering \cite{leis2021lifting}. Topological insulators are also resistant to conventional electrostatic confinement due to Klein tunneling \cite{beenakker2008colloquium, gutierrez2016klein, xie2017spintronic} which prevents the formation of localized states using electrostatic potentials alone. The introduction of magnetic fields, mass gaps, or physical boundaries are usually required to achieve confinement \cite{bahramy2012emergent}. In ultrathin TI flakes, however, the finite thickness itself defines physical boundaries that impose quantum confinement conditions, leading to the formation of discrete energy levels within the bulk bands. This thickness-induced confinement offers a promising avenue to tune the electronic structure in systems hosting topologically protected surface states, where hybridization between surface and interface states, including Volkov-Pankratov states \cite{lu2020dirac}, can also emerge due to quantum tunneling.

Thickness-dependent ARPES and band structure studies on ultrathin Bi\textsubscript{2}Se\textsubscript{3} \cite{zhang2010crossover} and Sb\textsubscript{2}Te\textsubscript{3} \cite{wang2010atomically} films show that between 2 and 6 QL, quantum tunneling couples opposite surface states, opening a small gap, which closes above 6 QL as gapless surface states emerge. This hybridization gap is extremely small ($\leq$ 0.1 eV), making its reliable detection under room-temperature STS conditions challenging due to combined thermal and instrumental broadening effects. Importantly, such a small hybridization gap has a negligible influence on the bulk-derived quantum well states (QWS), which are primarily governed by finite-thickness confinement. In molecular beam epitaxy (MBE)-grown TI films presence of QWSs has been observed in 2D limit for films having thickness between 2 nm to 8 nm \cite{jiang2012fermi, song2015probing}. STM and STS are ideally suited for probing quantum confinement at the atomic scale as these techniques enable the direct visualization of QWSs, allowing measurements of energy quantization, quasiparticle lifetimes, and scattering processes in real space. While STM/STS investigations have successfully characterized QWS in epitaxially grown TI thin films \cite{wang2010atomically, alpichshev2010stm}, they remain largely unexplored in exfoliated TI. The stability of topological surface states in ambient conditions is crucial for applications. Studies on Bi\textsubscript{2}Se\textsubscript{3} and Bi\textsubscript{2}Te\textsubscript{3} single crystals reveal that topological order remains robust, despite notably surface changes such as formation of 2D QWSs \cite{chen2012robustness}. Exfoliated flakes, free from substrate strain and consequent growth defects, offer a pristine platform to study confinement effects and intrinsic surface electronic structure. Open questions remain about how confinement manifests in mechanically exfoliated TI flakes, especially the robustness of QWS and their thickness dependent energy evolution. Additionally, the roles of surface quality, interlayer coupling, and substrate effects in shaping the electronic structure require further study.

(Bi\textsubscript{1-x}Sb\textsubscript{x})\textsubscript{2}Te\textsubscript{3} alloys have played a pivotal role \cite{hsieh2008topological} in the discovery of topological surface states by suppressing bulk conductivity and enabling clearer observation of surface-dominated phenomena. (Bi\textsubscript{0.1}Sb\textsubscript{0.9})\textsubscript{2}Te\textsubscript{3}(0001) is chosen for its well-defined surface states with a Dirac point in the vicinity of Fermi level \cite{sinha2025situ}, ideal for probing Dirac fermion-driven electronic behavior \cite{he2015substitution}. Its layered crystal structure, held together by weak van der Waals forces, allows mechanical exfoliation into ultrathin flakes comprising only a few quintuple layers, making it suitable for studying two-dimensional topological effects.

\begin{figure*}
   \centering
       \includegraphics[width=\textwidth]{"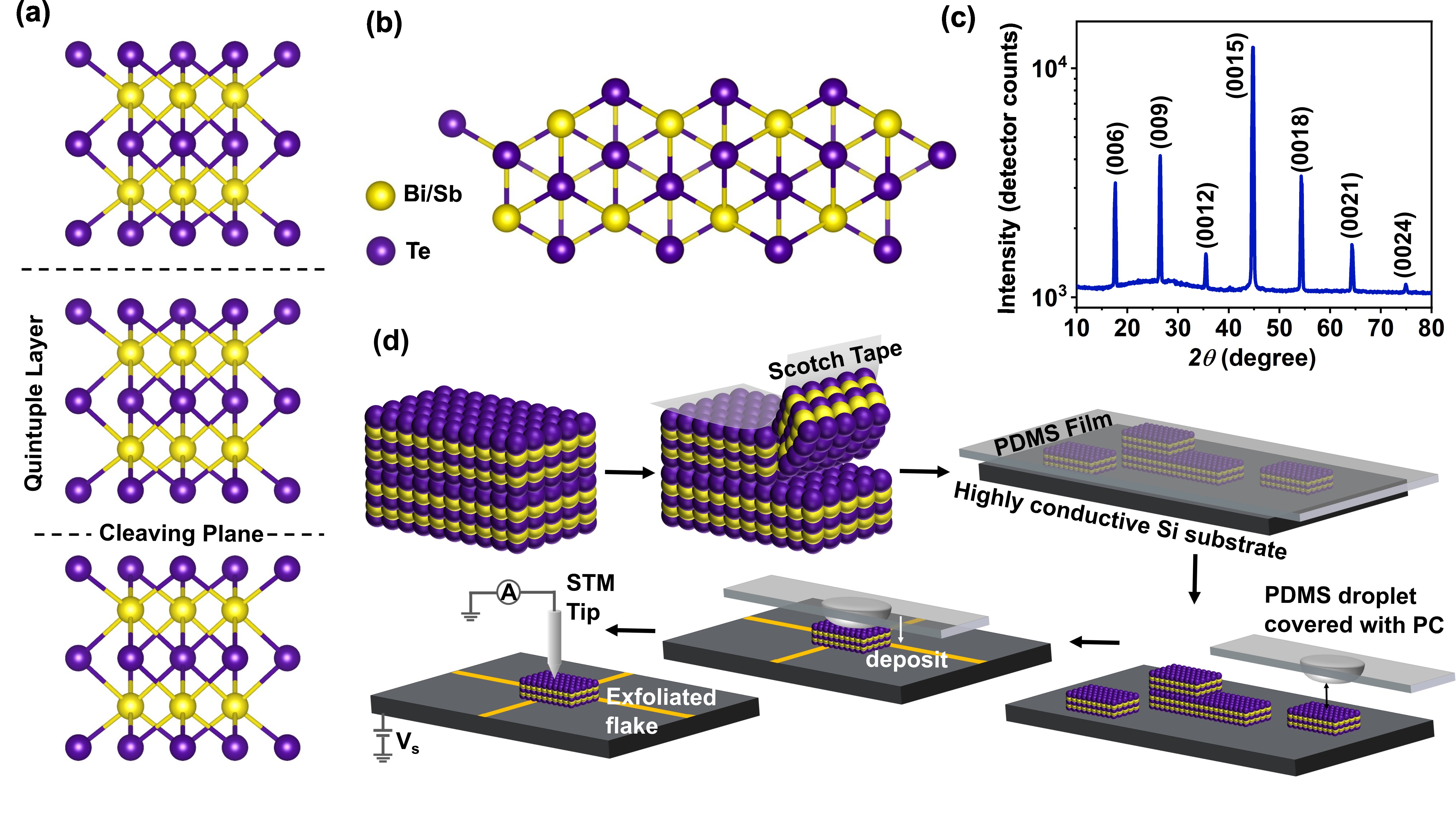"}
    \caption{(a) Schematic of layered atomic structure of (Bi\textsubscript{0.1}Sb\textsubscript{0.9})\textsubscript{2}Te\textsubscript{3} portraying quintuple layer sequence and the cleaving planes, single unit cell consists of three QLs. (b) Top view of the crystal structure. (c) X-ray diffraction spectrum of the as-grown (Bi\textsubscript{0.1}Sb\textsubscript{0.9})\textsubscript{2}Te\textsubscript{3} (0001) single crystal plotted on a logarithmic intensity scale. The vertical axis represents the raw detector counts (no normalization was applied), while the horizontal axis denotes the diffraction angle ($2\theta$). (d) Schematic of the mechanical exfoliation and dry-transfer technique for fabrication of ultra-thin (Bi\textsubscript{0.1}Sb\textsubscript{0.9})\textsubscript{2}Te\textsubscript{3} flakes: exfoliation from bulk crystal onto polydimethylsiloxane (PDMS) film to avoid adhesive contamination, transfer of flakes onto conductive Si substrate using a 2D transfer system with controlled heating for conformal contact, secondary transfer of selected thickness single flake onto another conductive Si substrate having Au cross as reference markers, using polycarbonate (PC)-coated PDMS handles for STM measurements, followed by PC removal with toluene.}
    \label{Figure1}
\end{figure*}

In this study, we investigated mechanically exfoliated (Bi\textsubscript{0.1}Sb\textsubscript{0.9})\textsubscript{2}Te\textsubscript{3} flakes deposited on a silicon substrates, with thicknesses ranging from approximately 3 quintuple layers (QL) or 3 nm (one unit cell) to 96 nm, as determined by atomic force microscopy (AFM). A combination of AFM and optical microscopy was employed to facilitate an initial quantitative analysis correlating flake thickness with their observed optical contrast. The distinct Raman signature of the material, though influenced by the sample volume and instrument settings \cite{van2023thickness}, was systematically analyzed under consistent conditions to serve as a reliable thickness indicator. This mechanical exfoliation approach provides a substrate-independent, time-efficient alternative to liquid-phase exfoliation \cite{sun2014preparation}, while still yielding high-quality flakes comparable to those produced by layer-controlled MBE \cite{jiang2012fermi, song2015probing}. STS measurements were performed to examine the dependence of the electronic band structure on number of QL, revealing signatures of QWS within the bulk valence band for exfoliated flakes with thicknesses $\leq$ 10 QL, consistent with previous reports on epitaxial films \cite{jiang2012fermi, song2015probing}. The energies of the observed QWS were analyzed as a function of the bulk band dispersion along the $\Gamma$–L direction within the –0.1 to –0.6 eV range relative to the Fermi level (E\textsubscript{F}), employing the phase accumulation model (PAM) \cite{becker2010scattering, echenique1978existence, smith1985phase}. The extracted electronic structure characteristics match well with the density functional theory (DFT) calculations. Furthermore, the interlayer vacuum regions exhibit charge depletion, while individual quantum layers show corresponding charge accumulation. This redistribution of charge reflects a polarization-like interlayer coupling predominantly mediated by van der Waals interactions. Spatially resolved spectroscopy around surface defects demonstrates that the QWS in two-dimensional TI remains resilient to defect-induced scattering, confirming their origin as confined Dirac electrons. This work provides the first direct STM/STS evidence of quantum well states in mechanically exfoliated (Bi\textsubscript{0.1}Sb\textsubscript{0.9})\textsubscript{2}Te\textsubscript{3} flakes, demonstrating that such quantized states persist independently of epitaxial strain or substrate coupling and are intrinsic to finite-thickness confinement in topological insulators.

\section{Experimental Methods}

Samples with varying layer numbers were prepared on a conductive silicon substrate using a mechanical exfoliation technique inside an argon-filled  glovebox (O\textsubscript{2} and H\textsubscript{2}O levels $\leq$ 0.1 ppm). Bulk single crystals of the topological insulator (Bi\textsubscript{0.1}Sb\textsubscript{0.9})\textsubscript{2}Te\textsubscript{3} were grown by self-flux method \cite{sinha2025situ, sinha2025magnetic}, and the layers were subsequently exfoliated using adhesive scotch tape. They were transferred to a polydimethylsiloxane (PDMS) film to avoid transfer of glue from the tape to the substrate \cite{huang2020universal}. All exfoliation, flake transfer and alignment steps-including optical inspection within the transfer system, were performed entirely inside the glovebox to ensure clean and oxidation-free interfaces. Before flake deposition, the conductive Si substrates were cleaned outside the glovebox sequentially with isopropanol (IPA), acetone, IPA and deionized water to remove organic contaminants and then dried using a nitrogen gas. Since this step does not involve the exfoliated flakes, it can be safely performed in ambient conditions without compromising surface quality. The cleaned substrates were then immediately introduced into the glovebox through an antechamber to minimize any atmospheric exposure before flake transfer. No buffered oxide etchant (BOE) treatment was applied to maintain a stable and inert surface for flake transfer. Using an \textit{in-situ} and automatic 2D transfer system (HQ Graphene), the PDMS film was gently brought into proximity with the conductive Si substrate, which was then heated to 100 \textsuperscript{o}C to promote thermal expansion and establish conformal contact. After 5 minutes, the temperature was reduced to 60 \textsuperscript{o}C, and the PDMS was slowly peeled off, leaving the (Bi\textsubscript{0.1}Sb\textsubscript{0.9})\textsubscript{2}Te\textsubscript{3} flakes adhered to the substrate. The spatial distribution of these flakes and their locations on different samples were initially estimated with the help of optical microscope of the transfer system. The thickness of these flakes was measured by atomic force microscopy (AFM) (Asylum Research, Oxford Instrument). The AFM measurements were performed using the tapping mode. In order to investigate the thickness-dependent Raman scattering a laser excitation at 532 nm was used. Raman spectra were collected on specific flakes of desired thickness that were already characterized using optical microscopy and AFM. Both the AFM and Raman measurements were performed outside the glovebox under ambient condition. After assigning thicknesses to all flakes, single flakes of selected thickness were transferred onto the center of another highly conductive silicon substrate, one-by-one, suitable for high-resolution STM measurements. Highly conductive Si (111) substrates ($\rho$ = 0.001 ohm cm) were chosen to ensure stable tunneling conditions during STM and STS measurements, as insulating SiO\textsubscript{2} surfaces often lead to charging effects and unstable current detection. For the transfer, hemispherical PDMS handles coated with a thin polycarbonate (PC) layer were used \cite{zomer2014fast}: the target flake was first picked up at 100 \textsuperscript{o}C by contacting the PC surface. Subsequently it was aligned and deposited onto the silicon substrate by pressing at 190 \textsuperscript{o}C, allowing the melted PC to release the flake Fig. \ref{Figure1}d. After the flake transfer, the samples were annealed in a 99.999\% pure (5N) argon atmosphere at 180 \textsuperscript{o}C for 2h to remove polymer residues and ensure surface cleanliness prior to STM imaging.The argon flow was additionally purified through a moisture and oxygen trap to prevent oxidation. This annealing temperature was found to be optimal: lower temperatures resulted in incomplete removal of polycarbonate residues, while higher temperatures ($\ge$ 200 \textsuperscript{o}C) occasionally caused partial flake degradation or delamination. Within this range, no observable change in surface morphology or doping was detected, confirming that the process effectively cleaned the surface without altering intrinsic properties. To enable unambiguous relocation in the STM, each selected flake was deterministically placed at the center of a pre-patterned Au cross deposited on the Si chip using a stainless-steel shadow mask. The Au features served as macroscopic reference markers visible to the naked eye and produced clear step edges in coarse STM scans. After mounting the chip, the Au cross was located first by stage coordinates, then a short low-resolution scan was performed to find the Au–Si step and translated to the flake positioned at the cross center. Thickness-dependent scanning tunneling microscopy (STM) measurements were conducted at room temperature using a Quazar nanoReV STM system \cite{sinha2025occurrence} equipped with an \textit{insitu} home-built integrated active vibration cancellation system \cite{pabbi2018anita}. The STM tips were fabricated from polycrystalline PtIr or W wires, which were chemically etched and subsequently annealed at high temperatures to ensure optimal performance. STS data was obtained using a lock-in technique to record the differential tunneling conductance (dI/dU) by adding an AC modulation voltage (frequency = 1.9 kHz, $V_{\text{rms}}$ = 10-50 mV)  to the bias voltage while incrementally varying the applied bias U. STM and STS data were digitally post-processed using WSxM software \cite{he2018exchange} and MATLAB. Post-processing included background plane subtraction and line-by-line flattening to remove scanner-induced tilt and thermal drift artifacts. Gentle smoothing filters were applied solely to reduce high-frequency instrumental noise without altering any intrinsic topographic or spectroscopic features. For STS measurements, multiple dI/dU spectra acquired under identical tunneling conditions were averaged to enhance the signal-to-noise ratio. The resulting spectra were then normalized with respect to the setpoint current and bias voltage to ensure quantitative consistency and enable reliable comparison across different sample thicknesses.

\begin{figure*}
   \centering
       \includegraphics[width=0.9\textwidth]{"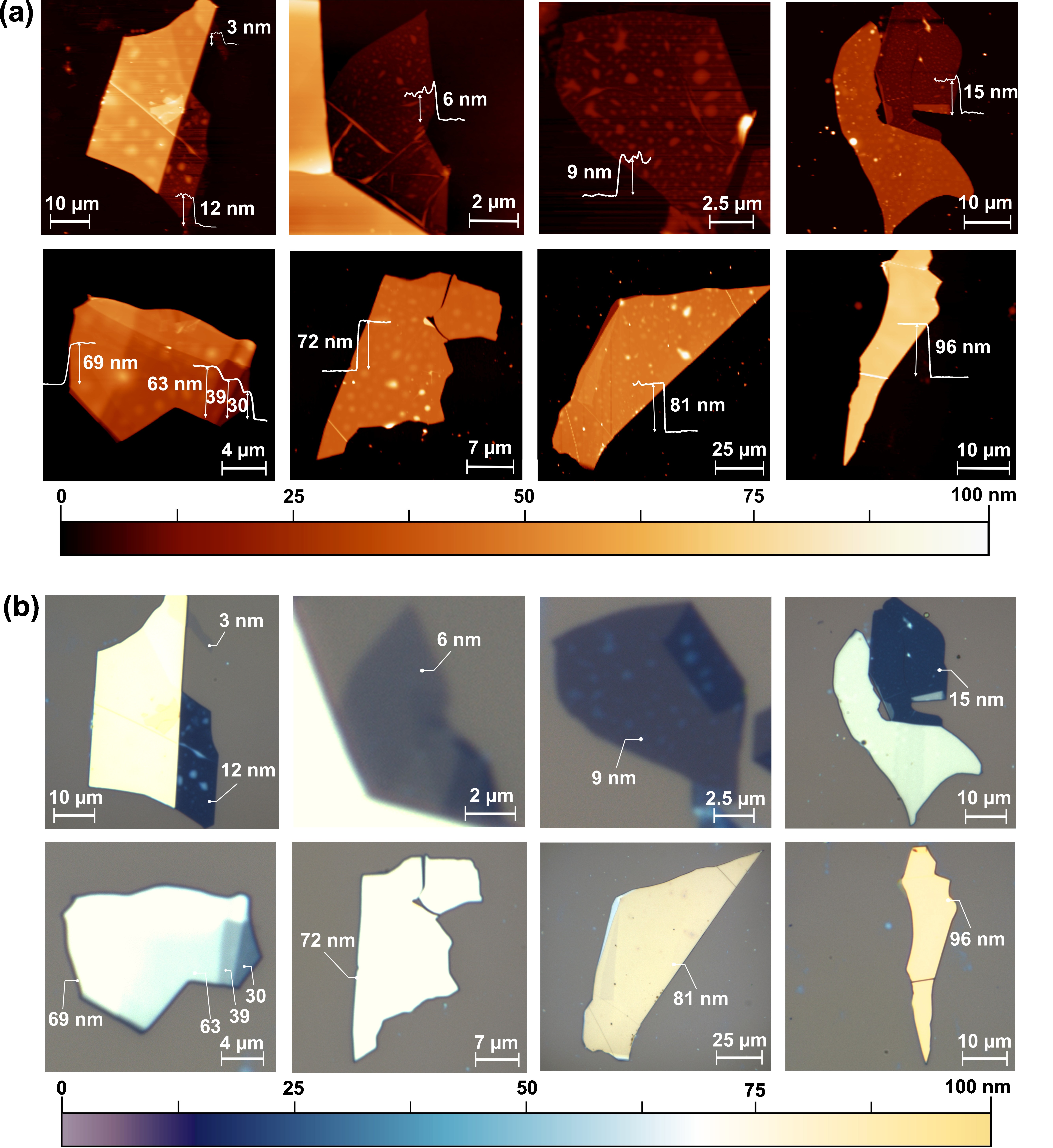"}
    \caption{Quintuple layer (QL) number-dependent apparent colors of exfoliated (Bi\textsubscript{0.1}Sb\textsubscript{0.9})\textsubscript{2}Te\textsubscript{3} flakes transferred to conductive Si-substrates: (a) AFM measurements of TI flakes with thicknesses ranging from 3 nm ($\sim$3 QL) and its integral multiples to 96 nm (96 QL). (b) Optical microscopy images of the corresponding flakes. }
    \label{Figure2}
\end{figure*}

\section{Computational Methods}

Density functional theory (DFT)~\cite{hohenberg1964inhomogeneous,kohn1965self} calculations were performed using the plane-wave pseudopotential method as implemented in the Vienna \textit{Ab initio} Simulation Package (VASP)~\cite{kresse1996efficiency,kresse1996efficient}. For all the theoretically modeled systems viz. bulk and layered ones, the structural relaxation was employed using the generalized gradient approximation (GGA) with the Perdew–Burke–Ernzerhof (PBE) exchange–correlation functional~\cite{perdew1996generalized}. To accurately capture weak interlayer interactions in layered structures, we have incorporated van der Waals corrections using the DFT-D3 scheme by Grimme~\cite{grimme2010consistent}. All structures were relaxed until the total energy was converged below 0.01~meV and atomic forces were less than $10^{-5}$~eV/\AA. For bulk Sb$_2$Te$_3$, the Brillouin zone was sampled using a $13 \times 13 \times 13$ Monkhorst–Pack $k$-point mesh. The kinetic energy cutoff for the plane-wave basis set was set to 600~eV. Spin–orbit coupling (SOC) was explicitly included in the self-consistent calculations for the electronic band structure and density of states due to the presence of heavy elements. Note that the majority of our theoretical calculations were performed on pristine Sb$_2$Te$_3$, whereas the experimental samples involved 10\% Bi doping. Therefore, it may appear that this could lead to some difference in the respective results viz. theory vis-à-vis experiment. However, this Bi doping was essentially done to modulate the chemical potential of the system and has no significant role near the conduction and valence band edges. To validate this, a representative calculation on a 2~QL (Bi$_{0.1}$Sb$_{0.9}$)$_2$Te$_3$ system was performed, confirming negligible Bi contribution near the band edges (see Supplemental Material \cite{SM}, Fig.~5). In our theoretical calculations, the primary aim is to analyze only the bands near the conduction and valence band edges. Inclusion of Bi atoms as dopants into the supercell would result in a much larger cell with a higher number of states (in addition to the quantum states of interest) that are mostly irrelevant for our analysis. Therefore, since the low-energy band dispersion remains unaltered and the quantum states of interest are unaffected by Bi doping, we preferred to analyze the smaller pristine Sb$_2$Te$_3$ supercell for the rest of our calculations. This not only reduces the computational cost but also provides an accurate description of the essential electronic structure of the experimentally synthesized doped compound to explore its low-energy band dispersion.

\section{Results and Discussion}

In order to evaluate the suitability of (Bi\textsubscript{0.1}Sb\textsubscript{0.9})\textsubscript{2}Te\textsubscript{3} for the experiments, the structural and crystallographic properties of the bulk material were examined. The crystal structure of this ternary TI is rhombohedral (space group $R\overline{3}m$, No. 166) and composed of van der Waals (vdW)-bonded quintuple layers (QLs), within which 10\% of Sb sites are randomly substituted by Bi atoms \cite{akiyama2018shubnikov, chong2020severe}. Cleaving typically occurs between adjacent QLs, yielding Te-terminated surfaces characterized by atomically flat terraces separated by step heights corresponding to a single QL (9.5–11 \AA). The topmost Te atoms are arranged in a hexagonal lattice with an in-plane atomic spacing of ~4.3 \AA \cite{chong2020severe}. Figures \ref{Figure1}(a–b) illustrates the side and top views of the (Bi\textsubscript{0.1}Sb\textsubscript{0.9})\textsubscript{2}Te\textsubscript{3} unit cell, highlighting the layered structure and surface symmetry. Structural integrity and crystallinity were further verified by \textit{ex-situ} X-ray diffraction (XRD) measurements, performed on a cleaved 0.4 mm thick crystal. As shown in Fig. \ref{Figure1}c, the XRD pattern displays sharp diffraction peaks indexed as (00n) reflections \cite{lanius2016topography, fan2006characterization}, confirming the high quality and preferred orientation along the c-axis of the single crystal. Figure \ref{Figure1}d illustrates the mechanical exfoliation process employed to obtain thin flakes of (Bi\textsubscript{0.1}Sb\textsubscript{0.9})\textsubscript{2}Te\textsubscript{3} using the scotch tape method. This technique, as detailed in the methodology section, enables the isolation of ultrathin layers by cleaving the crystal along its natural van der Waals planes.

\begin{figure*}
   \centering
       \includegraphics[width=\textwidth]{"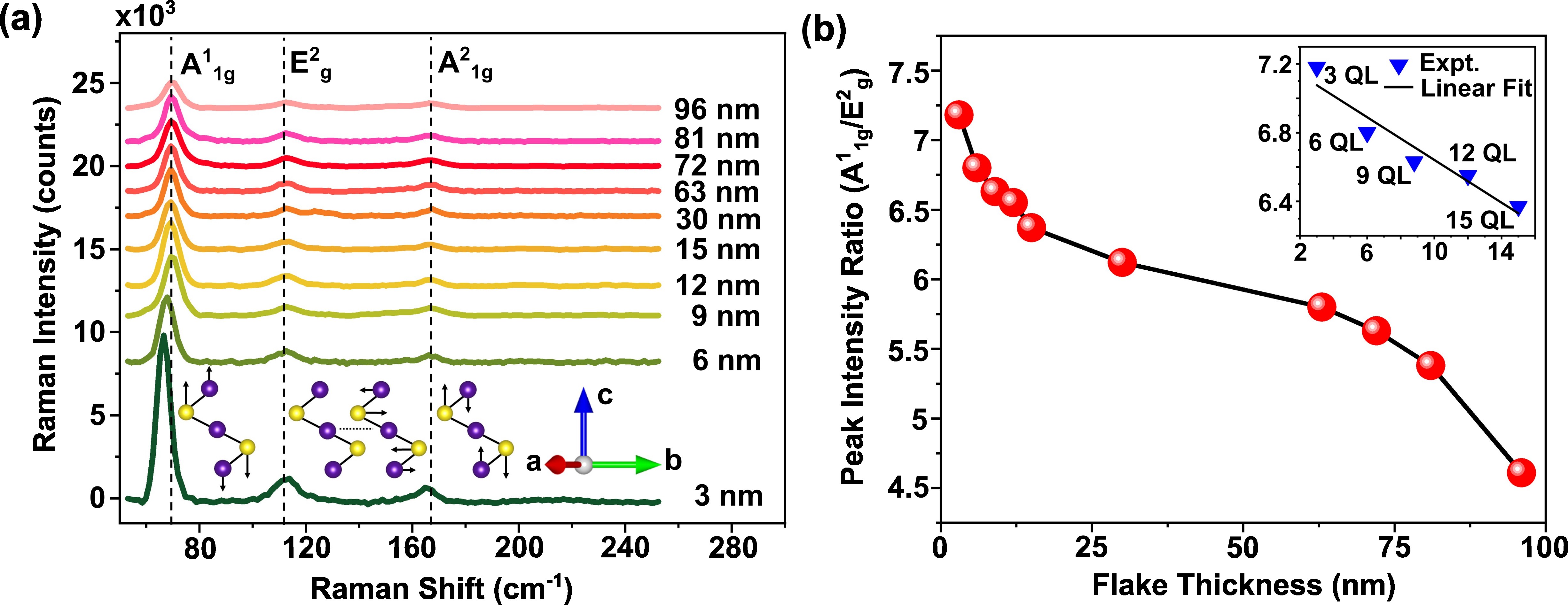"}
    \caption{(a) Raman spectra of exfoliated (Bi\textsubscript{0.1}Sb\textsubscript{0.9})\textsubscript{2}Te\textsubscript{3} flakes with different thicknesses showing the evolution of characteristic phonon modes with the inset illustrating their vibrational schematics and corresponding crystallographic directions. (b) Thickness dependence of the Raman peak intensity ratio $A^1_{1g}$ to $E^2_g$ for flakes ranging from a few QLs to bulk, (inset) Peak intensity ratio vs flake thickness for flakes  $\leq$ 15 QL highlighting their linear dependency in low thickness regime.}
    \label{Figure3}
\end{figure*}

Following exfoliation, the resulting flakes were characterized to assess their thickness and surface morphology. The spatial distribution and the thickness of the exfoliated (Bi\textsubscript{0.1}Sb\textsubscript{0.9})\textsubscript{2}Te\textsubscript{3} flakes were examined using a combination of optical microscope and AFM. Figure \ref{Figure2}a shows the AFM topography of the selected TI flakes with different thicknesses and Fig. \ref{Figure2}b portrays corresponding optical microscopy images. A direct comparison of AFM and optical images enabled the creation of a color chart \cite{zhang2021optical, puebla2020optical} that correlates the apparent colors of the (Bi\textsubscript{0.1}Sb\textsubscript{0.9})\textsubscript{2}Te\textsubscript{3} flakes on Si substrate with their corresponding thickness, with an uncertainty of $\pm$ 2 nm. The precise heights of the flakes were determined by analyzing line profiles along their edges in the AFM images. The thinnest flake exhibits a height of 3 nm, corresponding to approximately three quintuple layers (QLs) or a single unit cell of the material. The remaining flakes display thicknesses in integer multiples of QL heights, extending up to 96 nm. This behavior originates from the inherent cleavage mechanism between QLs, which are weakly bonded by van der Waals (vdW) interactions \cite{kong2010few}, \cite{teweldebrhan2010exfoliation}. However, the apparent preference for multiples of three QL arises from the limited vertical resolution of AFM under ambient conditions, where adsorbates and tip convolution can  obscure fine height variations on the order of a single QL ($\sim$1 nm) and merge closely spaced steps into averaged thicknesses such as 3 QL, 6 QL, or 9 QL. High-resolution STM measurements under UHV later revealed continuous layer-by-layer thickness variations (3–10 QL), with discrete steps corresponding to 3 QL, 4 QL, 5 QL etc. confirming that the AFM observation reflects an instrumental artifact rather than an intrinsic exfoliation preference. While some flakes exhibit a uniform surface, others demonstrate variations in height distribution. The flakes have lateral sizes ranging from few microns to tens of microns, which renders them well-suited for utilization in device fabrication and various industrial applications.

To establish a comprehensive trend of Raman scattering peaks as a function of exfoliated flake thickness, Raman spectra were recorded for flakes of varying thicknesses (Fig. \ref{Figure3}a), all measured at room temperature using a 532-nm excitation laser. The spectra were collected at precisely identified locations on the flakes, as determined by AFM measurements, ensuring accurate thickness identification. Within the scanned frequency range, three prominent peaks were observed. Based on the Raman selection rules outlined in Refs. \cite{richter1977raman, nemov2018specific}, these peaks, located at 69 cm\textsuperscript{-1}, 112 cm\textsuperscript{-1} and 170 cm\textsuperscript{-1} are assigned to \({A}_{1g}^1\), \({E}_{g}^2\) and \({A}_{1g}^2\) modes respectively. The \({A}_{1g}\) modes correspond to out-of-plane vibrational motion, whereas the  \({E}_{g}\) mode corresponds to in-plane vibrations. With increasing thickness (number of QLs) systematic reduction in the intensity of all peaks is observed, which is most pronounced in the low-thickness (2D) regime. In this regime, the \({A}_{1g}^1\) peak exhibits a red shift of 2.7 cm\textsuperscript{-1} for 3 QL flake and 1.4 cm\textsuperscript{-1} for 6 QL flake, indicating thickness-dependent vibrational behavior. The observed trend can be attributed to phonon softening, with the \({A}_{1g}^1\) mode exhibiting a greater sensitivity to thickness. As the number of layers decreases, the interlayer van der Waals forces weaken, leading to a reduction in the effective restoring forces acting on the atoms involved in out-of-plane vibrations \cite{zhang2011raman, wang2013situ}. In contrast, the \({E}_{g}^2\) mode undergoes a minimal blue shift when transitioning from the 3D to 2D regime \cite{zhao2014interlayer, singh2021study}. This shift can be explained by lattice contraction, which reduces interlayer pressure and increases the restoring forces acting on in-plane vibrations.

\begin{figure*}
   \centering
       \includegraphics[width=\textwidth]{"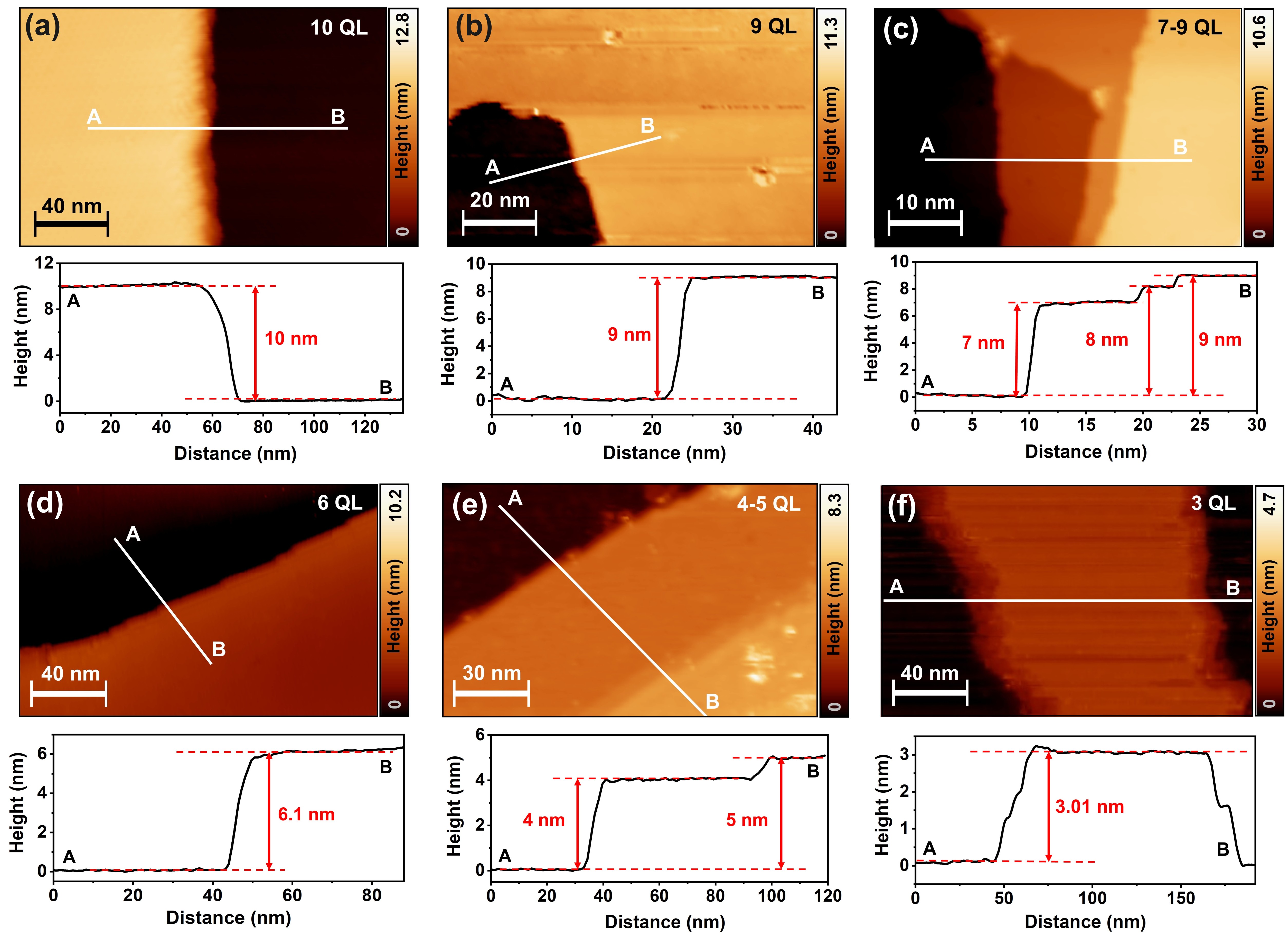"}
    \caption{(a–f) STM topographic images of exfoliated flakes with different thicknesses, ranging from 3 QL to 10 QL. All the topographic scans acquired at constant current mode with setpoint U = 500 - 800 mV, I = 100 pA (the colorbar represents relative height). The corresponding line profiles (shown below each image) show the step heights of the flakes relative to the substrate.}
    \label{Figure4}
\end{figure*}

\begin{table*}[t]
\caption{Intensity ratios of \({A}_{1g}^1\) to \({E}_{g}^2\) Raman peak at different flake thicknesses.}
\begin{ruledtabular}
\begin{tabular}{ccccccccccc}
Thickness (nm) & 3.0 & 6.1 & 8.8 & 12 & 15 & 30 & 63 & 72 & 81 & 96 \\
\hline
Peak intensity ratio (${A}_{1g}^1/{E}_{g}^2$) & 7.18 & 6.80 & 6.63 & 6.55 & 6.37 & 6.12 & 5.80 & 5.63 & 5.38 & 4.61 \\
\end{tabular}
\end{ruledtabular}
\end{table*}

The variation in the relative intensities of these two modes provides additional insight into the thickness dependence. As depicted in Fig. \ref{Figure3}b, the Peak Intensity Ratio (${A}_{1g}^1/{E}_{g}^2$), decreases with increasing flake thickness. The ratio exhibits an approximately linear trend in the thin-flake regime (3–15 nm), as highlighted in the inset, while for thicker samples, the decrease becomes gradual and deviates from strict linearity. This behavior originates not from a continued increase in structural restriction but from the distinct scaling of the two phonon modes. While the absolute intensities of both modes tend to saturate for bulk-like flakes, their relative attenuation rates differ slightly. Additionally, optical interference and the limited optical penetration depth in thicker flakes alter the effective Raman collection efficiency for out-of-plane and in-plane modes differently. Since the $A_{1g}$ mode is more sensitive to surface and c-axis vibrations, it experiences stronger attenuation with increasing thickness compared to the $E_{g}$ mode, resulting in a slow but continued decline in the intensity ratio even in the thick-flake regime. A similar trend has been reported by Shahil et al. \cite{shahil2012micro} for layered chalcogenide systems (Bi\textsubscript{2}Te\textsubscript{3}), where the out-of-plane to in-plane mode ratio continues to decrease up to the bulk limit. Table 1 presents the peak intensity ratios (${A}_{1g}^1/{E}_{g}^2$) for flakes with different thicknesses. Such a trend offers a practical approach for reliably determining the thickness of arbitrarily chosen flakes.

\begin{figure*}
   \centering      
        \includegraphics[width=\textwidth]{"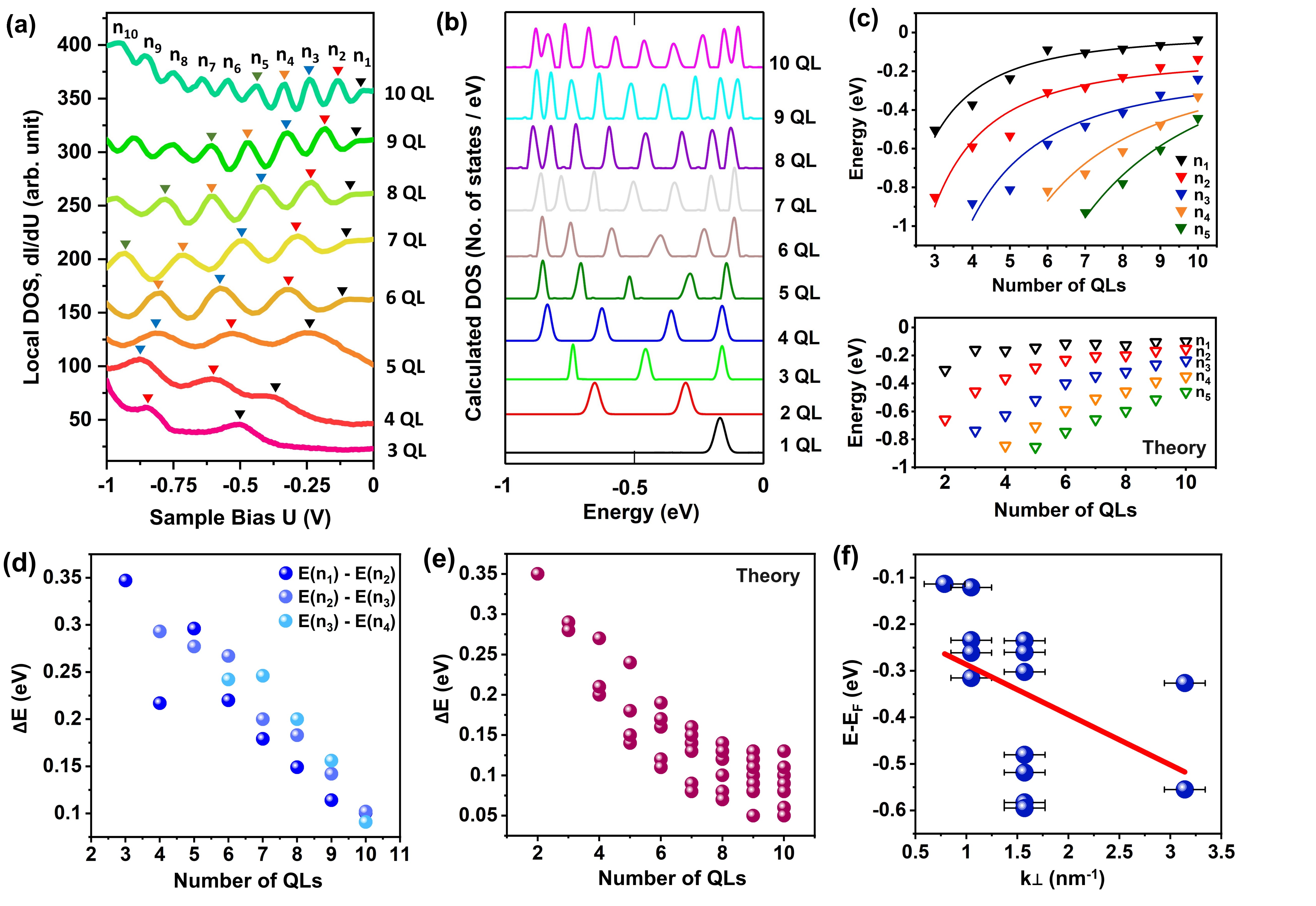"}
    \caption{(a) A series of LDOS spectra (vertically shifted) acquired on flakes of 3 QL to 10 QL height portray the quantization of electrons in 2D limits, and the arrows indicate the position of the corresponding QWS peaks of each layer. (b) Projected density of states (PDOS) at the $\Gamma$ point, resolved by QL number from 1 to 10, showing progressive peak sharpening and energy shifts. (c) Quantized energies of the QWSs as a function of flake thickness for different n-values (n=1-5, as shown with different colored triangles). The experimental data (top) are fitted using Nearly free electron model. (d) Energy spacing between adjacent dI/dU peaks as a function of the number of layers in 3-10 QLs (Bi\textsubscript{0.1}Sb\textsubscript{0.9})\textsubscript{2}Te\textsubscript{3}. Different color markers represent the energy separations between successive quantum well states ($n_1-n_2$, $n_2-n_3$, and $n_3-n_4$). (e) Subband energy splitting at $\Gamma$ as a function of the number of QLs, indicating quantum confinement and interlayer coupling. (f) QWS energies vs momentum plot giving $|v_g|= 1.6\times10^5$ m/s as calculated from the slope of the nearly linear fit.}
    \label{Figure5}
\end{figure*}

 To investigate their surface morphology and underlying electronic band structures, the exfoliated flakes were characterized using STM and STS. Figures \ref{Figure4}(a-f) present large-area STM topographic images of selected flakes transferred to highly conductive silicon substrates, captured at the interface between the exfoliated material and the underlying substrate. These flakes exhibit a range of thicknesses, varying from approximately 3.1 nm (3 QL) to 10 nm (10 QL). The associated line profiles, extracted along representative cross-sections, quantify the step heights at the flake–substrate boundary, confirming the vertical dimensions of the layers. Notably, while certain flakes display uniform thickness across the scanned region, others reveal a stratified morphology, consisting of multiple discrete terraces within a single flake. These variations in layer thickness across individual flakes suggest a degree of inhomogeneity that can arise during the mechanical exfoliation process. Such variations are key to investigate thickness-dependent quantum phenomena, especially the sensitivity of QWSs to local geometry and dimensionality. Atomically resolved STM images of the 10 QL flake (Fig. \ref{Figure7}d) and the 3 QL flake (Supplemental Material \cite{SM}, Fig. 1) further reveal the hexagonal lattice structure of Te-terminated surfaces.

Measurement of scanning tunneling spectroscopies over these mechanically exfoliated flakes approaching the two-dimensional limit reveals features that resemble QWSs. Thickness-dependent dI/dU spectra acquired from flakes with thicknesses between 3 and 10 quintuple layers (QLs) reveal a series of well-defined peaks corresponding to occupied electronic states. The regularly spaced oscillations within the valence band region are attributed to discrete QWSs arising from quantum confinement along the out-of-plane direction. These states reflect the formation of quantized energy levels within the potential well defined by the finite flake thickness. Systematic shifts in the QWS energies are evident as the thickness changes, indicative of enhanced confinement effects. As shown in Fig. \ref{Figure5}a, the 3 QL flake exhibits two distinct peaks at -0.503 eV and -0.85 eV, whereas the 4 QL flake displays three pronounced peaks at -0.373 eV, -0.59 eV and -0.883 eV. With increasing flake thickness, the number of observed peaks grows correspondingly, while the energy spacing between successive peaks decreases. In thicker flakes, a greater number of QWSs contribute to the tunneling spectra, leading to additional features at lower energies. This behavior is attributed to the hybridization of valence states from individual QLs, resulting in discrete subbands characteristic of the reduced-dimensional regime. The observed spectral peaks are thus interpreted as quantized two-dimensional QWSs, consistent with previous reports \cite{jiang2012fermi, song2015probing, wang2022real}. By 11 QL, the QWS are no longer observed (Supplemental Material \cite{SM}, Fig. 3). This may be attributed to the flake becoming sufficiently thick such that quantum confinement is strongly weakened and the subband spacing falls below the resolution of STS. In addition, as the thickness increases, the hybridization between the top and bottom surface states is largely suppressed, leading the electronic structure to approach bulk-like behavior without distinct quantization. No spectroscopic signature of one-dimensional edge states is observed at the step edges, which may require further exploration at low temperature.

\begin{figure*}
   \centering      
        \includegraphics[width=\textwidth]{"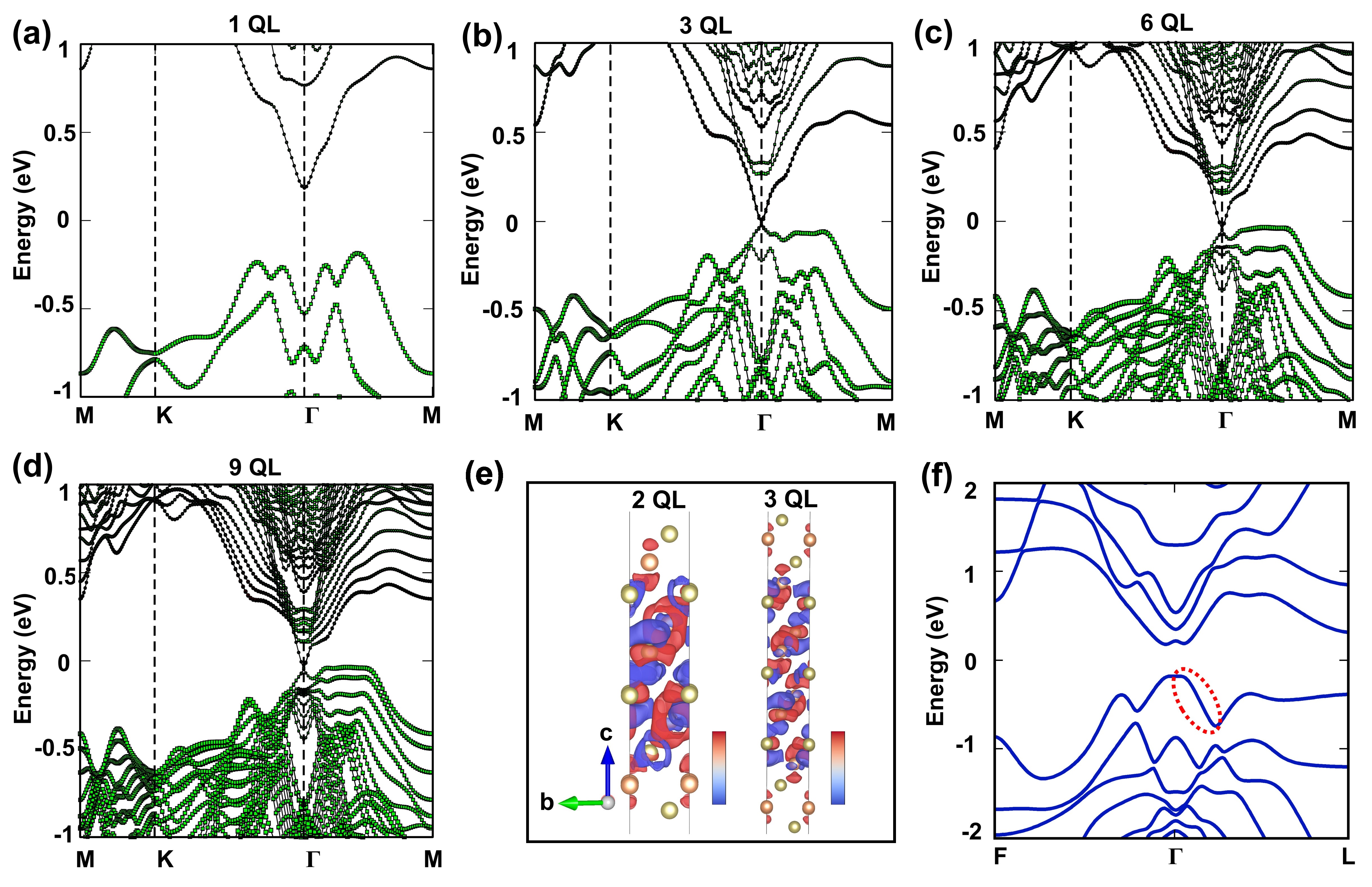"}
    \caption{(a–d) Electronic band structures of Sb$_2$Te$_3$ with 1~QL, 3~QL, 6~QL, and 9~QL thicknesses along the M–K–$\Gamma$–M path, illustrating the progressive subband splitting due to quantum confinement. The trend is consistent with the projected density of states [Fig.~5(b)], where increasing thickness leads to sharper subband features and reduced energy separation between quantized states. (e) Differential charge-density (CDD) plots for 2~QL (left) and 3~QL (right), showing charge accumulation (red) and depletion (blue) regions. The isosurface values are 9.83$\times$10$^{-5}$ and 1.35$\times$10$^{-4}$~e/\AA$^3$, with corresponding CDD ranges of (+7.16$\times$10$^{-4}$, -1.04$\times$10$^{-3}$) and (+7.60$\times$10$^{-4}$, -1.05$\times$10$^{-3}$)~e/\AA$^3$ for 2~QL and 3~QL, respectively. These plots reveal charge redistribution within quintuple layers and partial depletion in the interlayer region, consistent with the evolution of interlayer coupling and van der Waals interaction with increasing thickness. (f) Bulk Sb$_2$Te$_3$ band structure, exhibiting continuous bands without subband splitting, serving as a reference for the $\Gamma$–L dispersion extracted from the quantum well states. The red-dotted semicircle indicates the linear $E$–$k$ dispersion similar to the experimental result.}
    \label{Figure6}
\end{figure*}

The layer-resolved projected density of states (PDOS) at the $\Gamma$ point (Fig.~\ref{Figure5}b) exhibits a gradual sharpening and energy shifts of DOS peaks with increasing thickness from 1 to 10 QLs. This behavior reflects quantum confinement in thinner layers with more discrete and localized density of states and a clear increase in the same is visible near the band edges on increasing the QLs for thicker systems. The number of QWS peaks resolved in the experimental dI/dU spectra is slightly lower than that predicted by theory. This discrepancy can be attributed to experimental factors that limit the resolution of closely spaced subbands, primarily the finite energy resolution of STM arising from thermal broadening. Since all spectroscopy measurements were performed at room temperature, the combined effects of thermal broadening and finite AC modulation reduce the ability to distinguish very closely spaced QWS features at a given energy. On the theoretical side, the DFT-based slab calculations are performed under idealized conditions and do not account for temperature and tunneling effect. Despite this minor difference in the number of resolved QWSs, the experimentally observed QWS energies and their evolution with thickness exhibit excellent quantitative agreement with the calculated QW energy positions, supporting the quantum confinement interpretation.

\begin{figure*}
   \centering      
        \includegraphics[width=0.9\textwidth]{"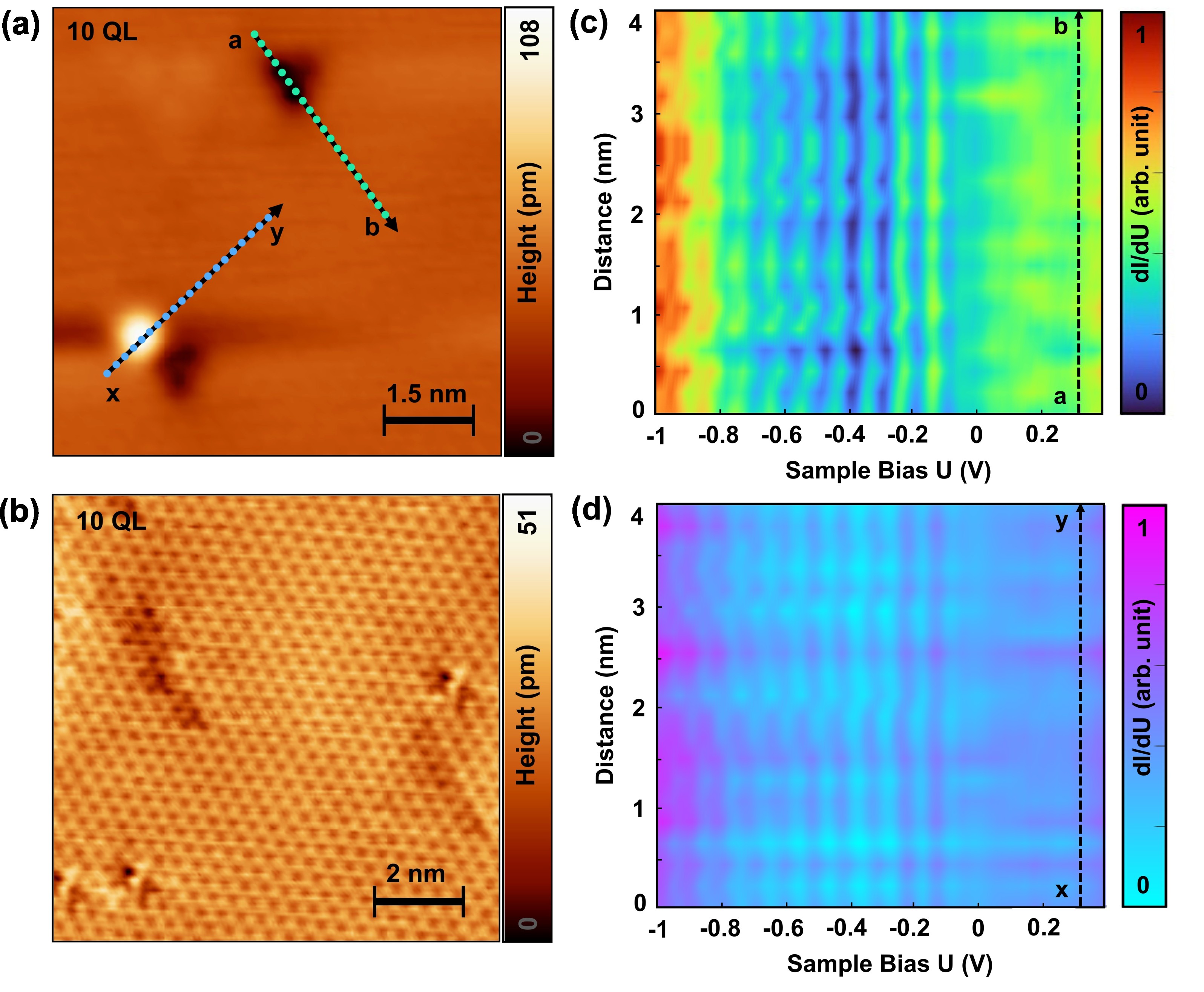"}
    \caption{(a) STM topograph (7.5 nm $\times$ 7.5 nm) of a 10 QL flake surface (U = 550 mV and I = 200 pA) showing a flat area with two standard TI defects, the triangular black depression (Te vacancies, V\textsubscript{Te}) and the bright protrusion (antisite defect, Te\textsubscript{Bi/Sb}). (b) Atomically resolved STM scan (10 nm $\times$ 10 nm) obtained on a nearby region on the same 10 QL flake (U = 300 mV, I = 500 pA) after confirming surface flatness and absence of large clusters. The image clearly reveals the characteristic hexagonal lattice arrangement and triangular defect motifs typical of the Te-terminated (Bi\textsubscript{0.1}Sb\textsubscript{0.9})\textsubscript{2}Te\textsubscript{3} surface, confirming the QWSs measurements were performed on the Te-terminated TI surface. (c) Two-dimensional projection of twenty position-dependent dI/dU spectra measured along a line of length 4 nm (marked by the black arrow from point a to b in panel (a)) crossing regions with and without the triangular defect; the blue dots represent the spectroscopic locations. (d) Similar two-dimensional projection of another twenty position-dependent dI/dU spectra acquired over the bright defect and away from it over a length scale of 4 nm (marked by the black arrow from point x to y in panel (a)); the purple dots show the spectroscopic locations of the twenty spectra. The uniformity and intrinsic nature of the QWSs across these regions demonstrate their robustness and confirm their origin from Dirac electron confinement within the ultrathin flakes, rather than defect-induced effects.} 
    \label{Figure7}
\end{figure*}

Fig. \ref{Figure5}c presents the layer-resolved quantum well state (QWS) energies as a function of flake thickness, extracted from the data shown in Fig. \ref{Figure5}a and Fig. \ref{Figure5}b, respectively. The experimental data have been quantitatively analyzed by fitting them with the Nearly Free Electron (NFE) model, where the geometrical confinement of electrons in the TI flakes is modeled as an infinite potential well, as proposed in \cite{zhong2013quantum}. This approximation imposes rigid boundary conditions, resulting in the quantization of energy levels according to the following expression:

\begin{equation}
E_0 + \frac{\hbar^2 \pi^2 n^2}{2 m^* d^2} = E_0 + \frac{A n^2}{d^2},
\end{equation}

where the flake thickness is defined as $d = aN$, with $N$ representing the number of quintuple layers and $a = 1$ nm being the thickness of the quintuple layer. In the fitting process, the quantum number $n$ is fixed for each curve, while $E_0$ and $A$ are treated as fitting parameters to account for deviations from ideal behavior and to accommodate material-specific effects such as non-parabolicity and effective mass variation.

Figures~\ref{Figure5}d and \ref{Figure5}e present the thickness dependence of the energy spacing between adjacent experimental peaks and the calculated subbands, including the energy splitting between the highest valence subbands at $\Gamma$, showing excellent quantitative agreement between experiment and theory. The splitting decreases with an increase in the QL number, consistent with a quantum well scenario in which interlayer coupling gives rise to quantized subbands along the out-of-plane direction. In Fig. \ref{Figure5}d, the different color markers denote the energy separations between successive quantum well states ($n_1-n_2$, $n_2-n_3$, and $n_3-n_4$). The energy spacing systematically decreases with thickness, confirming the weakening of confinement in thicker flakes. For higher QL numbers, such as 10 QL, the energy differences between $n_1-n_2$ (0.101 eV) and $n_2-n_3$ (0.102 eV) are nearly identical, leading to overlapping points that may appear as a single data point in the plot. This near-degeneracy signifies the gradual evolution from discrete quantized levels toward a quasi-continuous bulk band structure. From the energies of the QWSs as a function of flake thickness, the bulk-band dispersion along the along $\Gamma-L$ direction can be derived. For this purpose, the phase accumulation model (PAM) \cite{becker2010scattering} is employed, which is based on the Bohr–Sommerfeld quantization condition:

\begin{equation}
2k(E)(d + d_0) + \phi = 2\pi n,
\end{equation}

Here, $\phi = \phi_C + \phi_B$ denotes the total phase shift resulting from electron reflection at the crystal boundary ($\phi_C$) and at the vacuum interface ($\phi_B$), $k(E)$ represents the energy-dependent wave vector of an electron propagating along the $\Gamma$–$L$ direction in a flake of effective thickness $d + d_0$, where $d_0$ is an energy-independent offset and fit parameter. Importantly, $\phi$ is assumed to be independent of thickness and solely a function of energy.

Using this model, the wave vector $k(E)$ at a specific energy can be deduced from the difference in QWS quantum numbers at that energy. If two QWSs with identical energies $E = E_1 = E_2$ are observed in flakes of thicknesses $d_1$ and $d_2$, corresponding to quantum numbers $n_1$ and $n_2$ respectively, equation (2) simplifies to:

\begin{equation}
k(E) = \frac{\pi(n_2 - n_1)}{d_2 - d_1}
\end{equation}

Using Equation (3), $k(E)$ was computed for those QWS energy values where the difference $|E_1 - E_2|$ lay within a narrow tolerance ($\leq 50$ meV), ensuring the validity of the approximation. The resulting band dispersion along the $\Gamma$–$L$ direction is summarized in Fig. \ref{Figure5}f, plotted within an extended Brillouin zone framework. The data reveal a nearly linear dependence of energy $E$ on wave vector $k$ across the observed energy range. The slope of this dependence yields an approximate group velocity of $v_g = 1.6 \times 10^5$ m s$^{-1}$, reflecting the bulk-like electronic behavior of these TI flakes along the $\Gamma$–$L$ direction. This apparent linearity should not be regarded as a precise determination of the bulk band dispersion. STM/STS is inherently a real-space technique and does not directly provide momentum-resolved information, and the reconstructed $E$–$k$ trend relies on simplifying assumptions within the phase accumulation framework. Additional uncertainties arise from finite energy resolution, the experimental uncertainty in flake thickness ($\delta d$ $\approx$ 0.05 nm), and lifetime broadening of QWSs, contributing to an estimated error of $\delta k$ $\approx$ 0.2 nm$^{-1}$. Hence, the linear fit in Fig. 5f is a qualitative indication of nearly linear QWS evolution rather than a strict quantitative dispersion, consistent with the expected crossover from parabolic to linear conduction-band behavior near the $\Gamma$–$L$ direction in confined Dirac-like states, as discussed later.

Previously, the phase accumulation model has been extensively and successfully applied to the interpretation of image-potential and surface states in clean metallic systems, as well as to QWSs in noble-metals thin films. The present study demonstrates that this model can be effectively extended to analyze QWSs on the surfaces of TIs, thereby reinforcing its broader applicability in low-dimensional quantum systems.

Figures~\ref{Figure6} and Supplemental Material \cite{SM} Fig. 4 display the electronic band structures and interlayer interaction characteristics of Sb\textsubscript{2}Te\textsubscript{3} from monolayer to bulk, calculated along the high-symmetry path M–K–$\Gamma$–M. In the monolayer (1 QL, Fig.~\ref{Figure6}a). The highest valence band at $\Gamma$ (denoted ML–VB1@$\Gamma$) predominantly consists of Sb 5$p_z$ and Te 5$p_z$ orbitals. These out-of-plane orbitals are crucial for interlayer coupling in multilayer systems.

As additional layers are added, the electronic bands split due to van der Waals (vdW) interactions. In the bilayer configuration (2 QL, Supplemental Material \cite{SM} Fig.~4a), ML–VB1 gives rise to subbands (BL–VB1, BL–VB2, etc.) with lifted degeneracy, particularly at the $\Gamma$ point. For 6 QL and 9 QL [Figs.~\ref{Figure6}(c–d)], further subband evolution is evident as the system transition towards bulk-like behavior. Similar to the valence band evolution, the conduction band edge at the $\Gamma$ point also exhibits a noticeable lifting of degeneracy with increasing thickness. In the monolayer limit, the lowest conduction band (ML–CB1@$\Gamma$) is singly degenerate; however, as additional layers are introduced, this band splits into multiple subbands (e.g., BL–CB1, BL–CB2), reflecting interlayer coupling effects and the emergence of a quantum well-like potential in the out-of-plane direction. This behavior is particularly evident in the 3 QL and 6 QL band structures [Figs.~\ref{Figure6}(b–c)], where distinct conduction subbands appear at the $\Gamma$ point with progressively smaller energy separation. These conduction subband splittings mirror those observed in the valence band and further support the quasi-2D confinement picture. Consistently, the DFT band structures in Figs.~\ref{Figure6}(a–d) and Supplemental Material Fig.~4 reveal a progressive lifting of degeneracy at $\Gamma$ as the number of QLs increases. Experimentally, this manifests as an increasing number of QWS peaks and a reduction in energy spacing in dI/dV spectra with thickness, confirming that the observed STS peaks are the subbands generated by slab-thickness confinement.

Charge density difference (CDD) plots for 2 QL and 3 QL systems (Figs.~\ref{Figure6}e) illustrate charge depletion in the interlayer vacuum region and accumulation within individual QLs, consistent with a polarization-type interlayer interaction predominantly governed by van der Waals forces rather than covalent bonding. For comparison, the band structure of bulk Sb\textsubscript{2}Te\textsubscript{3} (Fig.~\ref{Figure6}f) confirms the closure of the subband splitting and emergence of continuous bulk bands. Furthermore, Fig.~\ref{Figure6}f serves as the reference for the $\Gamma$–L dispersion extracted from the QWSs via the phase accumulation model (Fig.~\ref{Figure5}f). The linear $E(k)$ obtained experimentally along $\Gamma$–L exhibits a slope $v_g$ consistent with the DFT band velocity near $\Gamma$ in Fig.~\ref{Figure6}f, validating the PAM-based reconstruction of the bulk valence dispersion from finite-thickness QWS energies. Agreement in curvature (velocity) between Fig.~\ref{Figure5}f and Fig.~\ref{Figure6}f confirms that the quantized states sample the bulk $\Gamma$–L band, and that the PAM provides an accurate boundary condition for the system slabs.

To further confirm the intrinsic origin and robustness of the quantum well states (QWSs), we performed detailed spatially resolved spectroscopic measurements on a 10 QL flake across regions containing characteristic topological insulator (TI) surface defects as well as pristine areas. The STM topograph (Fig. \ref{Figure7}a) reveals two standard defect types commonly observed on the Te-terminated surface of (Bi\textsubscript{0.1}Sb\textsubscript{0.9})\textsubscript{2}Te\textsubscript{3}: the triangular depression and the bright protrusion. To verify that the probed area indeed corresponds to the Te-terminated surface of (Bi\textsubscript{0.1}Sb\textsubscript{0.9})\textsubscript{2}Te\textsubscript{3}, we performed an additional high-resolution STM scan on a nearby region of the same flake where no large clusters were present, as such clusters can distort the atomic-scale imaging contrast. The resulting atomically resolved image (Fig. \ref{Figure7}b) clearly reveals the characteristic hexagonal lattice of Te atoms of the top layer and the triangular defect features, which are well-known signatures of this topological insulator surface \cite{sinha2025situ}. The imaging parameters for this scan  differ from those in Fig. \ref{Figure7}a to optimize stability and achieve atomic resolution. After confirming the atomic-scale surface structure, sets of position-dependent dI/dU spectra along two distinct 4 nm trajectories were acquired, as shown in panel a, marked by two black arrows: one over the triangular defect and away from it (from point a to b) and another crossing regions with the bright protrusion and without it (from point x to y). The two-dimensional projections of these two sets of twenty sequential dI/dU spectra along each trajectory reveal well-resolved and spatially periodic QWS features as portrayed in Fig. \ref{Figure7}c and Fig. \ref{Figure7}d respectively (the individual spectra are included in the Supplemental Material \cite{SM}). These spectra show no significant deviations in energy position or spectral intensity between the defective and defect-free regions. This invariance demonstrates that the quantized states persist uniformly across the surface, regardless of local structural imperfections. Such behavior highlights that the QWSs originate from electron confinement along the out-of-plane direction of the finite-thickness crystal, rather than being influenced or induced by local surface perturbations. The observation of identical QWSs across multiple defect types indicates that the confinement is dominated by the film geometry and intrinsic potential boundaries, rather than scattering from point defects or surface disorder. Moreover, the robustness of the QWSs against these localized imperfections underscores their intrinsic nature and resilience, consistent with the topological protection of the underlying electronic structure. The combination of atomically resolved imaging and spatially resolved spectroscopy thus provides direct evidence that the QWSs are intrinsic to the confined Dirac electronic states of the ultrathin topological insulator and remain unaffected by common surface defects or adsorbates. These results further reinforce that the confinement mechanism in exfoliated (Bi\textsubscript{0.1}Sb\textsubscript{0.9})\textsubscript{2}Te\textsubscript{3} arises from well-defined boundary conditions at the film interfaces and not from extrinsic defect-related effects, highlighting the material’s suitability for reliable exploration of quantum-confined topological phenomena.

\section{Conclusion}
Our study demonstrates the successful mechanical exfoliation of the three-dimensional topological insulator (Bi\textsubscript{0.1}Sb\textsubscript{0.9})\textsubscript{2}Te\textsubscript{3} into atomically clean ultra-thin layers down to three QLs. In flakes thinner than 10 QLs, we observe discrete QWSs of Dirac electrons within the bulk valence band. We establish a reliable methodology for the accurate determination of flake thickness using a combination of techniques. These include optical imaging, atomically resolved STM imaging, micro-Raman spectroscopy, and scanning tunneling spectroscopy. The observed well-defined QWS which arise from quantum confinement, are accurately captured by the phase accumulation model and supported by theoretical first principle calculations. Using high-resolution spectroscopy, we further demonstrate that the QWSs in topological insulators remain robust against local surface impurity and disorder. This work establishes exfoliated (Bi\textsubscript{0.1}Sb\textsubscript{0.9})\textsubscript{2}Te\textsubscript{3} as a scalable topological material platform for visualizing and engineering quantum-confined Dirac states. Our approach marks a crucial step towards tunable topological quantum devices, including exploration of novel Moiré topological phenomena.
 
\section{Acknowledgement}

S.S. acknowledges DST for providing INSPIRE fellowship [IF190537]. This work is funded by DST funded project ‘CONCEPT’ under nanomission program (DST/NM/QM-10/2019). S.S. and S.M. would like to thank Pankaj Singh Pokhriya, Saurav Sachin and Nazma Firdosh for the technical help during development of \textit{in-situ} automatic layer transfer setup. Authors also acknowledge the Department of Physics and Central Research Facility (CRF), IIT Delhi for providing various characterization supports. S.P. acknowledges PMRF, India, for the Research fellowship [Grant No. 1403227]. S. B. acknowledges financial support from SERB under a core research grant (grant no. CRG/2019/000647) to set up his High Performance Computing (HPC) facility “Veena” at IIT Delhi for computational resources.

\section{Data Availability}

The data are available from the authors upon reasonable request.

\bibliographystyle{apsrev4-2}
\bibliography{references}

\end{document}